# Solutions of the Schrödinger equation with Quarkonium potential to predict the mass-spectra of the heavy mesons via series expansion method

E. P. Inyang[1,2], E. P. Inyang[2], I. O. Akpan[2], and E. S. William[2]

**ABSTRACT**

In this study, a quarkonium potential is adopted as the quark-antiquark interaction potential for predicting the mass spectra of heavy mesons. We solved the radial Schrödinger equation analytically using the series expansion method and obtained the energy eigenvalues. The present results are applied for predicting the mass spectra of heavy mesons such as charmonium and bottomonium. The present potential provides satisfying results in comparison with experimental data and the work of other researchers with a maximum error of $0.058\,GeV$.

## INTRODUCTION

The study of heavy quarkonium system such as charmonium and bottomonium has an important role in understanding the quantitative tests of quantum chromodynamics (QCD) and the standard model (Mutuk,2018). This system can be studied within the Schrödinger equation (SE)( Kumar and Chand,2014). The solution of SE with spherically symmetric potential is one of the important problems in Physics and Chemistry. This is because it plays an important role in understanding of the properties of constituents' particles and dynamics of their interactions (Abu-Shady et al., 2019). The fundamental potential used in the studying quarkonium system is the Cornell potential, also known as Killingbeck potential with two important features of the strong interaction, namely, asymptotic freedom and quark confinement (Rai, and Rathaud 2015; Bettini,2018). The SE has been solved using various analytical methods such as, the asymptotic iteration method ( Kumar and Chand,2014; Ikot et al., 2020), Laplace transformation method (Abu-Shady et al., 2018), super symmetric quantum mechanics method (SUSQM) (Abu-Shady et al., 2021)., Nikiforov-Uvarov(NU) method(Inyang et al., 2021;Edet et al., 2020; Edet et al., 2020;Inyang, et al., 2021; Inyang et al.,Akpan et al.,2021; Edet et al., 2019;Inyang et al., 2021;William et al.,2020;Inyang et al.,2021; William et al.,2020;Inyang et al.,2020;Inyang et al., 2021;Edet et al., 2020;Omugbe et al., 2021),the series expansion method (Inyang et al., 2021) and so on. Most researchers have studied the mass spectra of heavy mesons with Cornell potential (Rani et al., 2018; Ciftci and Kisoglu 2018; Al-Jamel, 2019; Mansour and Gamal 2018; Al-Oun et al., 2015;Omuge et al.,2020). For instance, (Ali et al.2015) studied the energy spectra of mesons and hadronic interactions using Numerov's method. Their solutions were used to describe the phenomenological interactions between the charm-anticharm quarks via the model. The model accurately predicts the mass spectra of charmed quarkonium as an example of mesonic systems. Also, (Inyang et al.2021) obtained the Klein-Gordon equation solutions for the Yukawa potential using the NU method. The energy eigenvalues were obtained both in relativistic and non-relativistic regime. They applied the results to calculate heavy-meson masses of charmonium and bottomonium.

The quarkonium or heavy-quark potential model takes the form (Abu-Shady and Ikot 2019).

$$V(r,T) = \left(\frac{2a}{m_D^2(T)} - a\right)\frac{e^{-m_D(T)r}}{r} - \frac{2b}{m_D^2(T)r} + \frac{2b}{m_D(T)} - am_D(T) \qquad (1)$$

where $a$ and $b$ are potential strength parameters, $m_D(T)$ is Debye mass which is temperature dependent and vanishes as $T=0$. The aim of this work is to investigate the SE with the quarkonium potential in the framework of the series expansion method to predict the mass spectra of heavy quark- antiquark system. To the best of our knowledge this is the first time quarkonium potential is being studied with the aim of determining the mass spectra of heavy mesons.

The paper is organized as follows: In section 2, the bound state energy eigenvalues is calculated via series expansion method.

*Corresponding author. Email: etidophysics@gmail.com or einyang@noun.edu.ng
[1]Department of Physics, National Open University of Nigeria, Jabi, Abuja
[2]Theoretical Physics Group, Department of Physics, University of Calabar, P.M.B 1115 Calabar, Nigeria
.

In section 3, the results are discuss. In section 4, conclusion is presented.

## METHODS

**Bound state solutions of the Schrödinger equation with quarkonium potential**

We consider the radial SE of the form (Inyang *et al.*, 2021).

$$\frac{d^2R(r)}{dr^2}+\frac{2}{r}\frac{dR(r)}{dr}+\left[\frac{2\mu}{\hbar^2}(E_{nl}-V(r))-\frac{l(l+1)}{r^2}\right]R(r)=0 \quad (2)$$

where $l$ is angular quantum number taking the values 0,1,2,3,4…, $\mu$ is the reduced mass for the quarkonium particle, $r$ is the inter-nuclear separation and $E_{nl}$ denotes the energy eigenvalues of the system.

The expansion of the exponential terms in Eq. (1) (up to order three, in order to model the potential to interact in the quark-antiquark system) yields,

$$\frac{e^{-m_D(T)r}}{r}=\frac{1}{r}-m_D(T)+\frac{m_D^2(T)r}{2}-\frac{m_D^3(T)r^2}{6}+\ldots \quad (3)$$

Substituting Eq. (3) into Eq.(1) we have

$$V(r,T)=-\frac{\alpha_0}{r}+\alpha_1 r+\alpha_2 r^2+\alpha_3 \quad (4)$$

where

$$-\alpha_0=\frac{2a}{m_D^2(T)}-a-\frac{2b}{m_D^2(T)}, \quad \alpha_1=a-\frac{am_D^2(T)}{2}$$
$$\alpha_2=\frac{am_D^3(T)}{6}-\frac{am_D(T)}{3}, \quad \alpha_3=\frac{2b}{m_D(T)}-\frac{2a}{m_D(T)} \quad (5)$$

We substitute Eq.(4) into Eq.(2) and obtain

$$\frac{d^2R(r)}{dr^2}+\frac{2}{r}\frac{dR(r)}{dr}+\left[\varepsilon+\frac{G}{r}-Hr-Jr^2-\frac{L(L+1)}{r^2}\right]R(r)=0 \quad (6)$$

where

$$\varepsilon=\frac{2\mu}{\hbar^2}(E-\alpha_3),\ G=\frac{2\mu\alpha_0}{\hbar^2}$$
$$H=\frac{2\mu\alpha_1}{\hbar^2},\ J=\frac{2\mu\alpha_2}{\hbar^2} \quad (7)$$

$$L(L+1)=l(l+1) \quad (8)$$

From Eq. (8),

$$L=-\frac{1}{2}+\frac{1}{2}\sqrt{(2l+1)^2} \quad (9)$$

Now make an anzats wave function

$$R(r)=e^{-\alpha r^2-\beta r}F(r) \quad (10)$$

Where $\alpha$ and $\beta$ are positive constants whose values are to be determined in terms of potential parameters.

Differentiating Eq.(10) twice gave,

$$R'(r)=F'(r)e^{-\alpha r^2-\beta r}+F(r)(-2\alpha r-\beta)e^{-\alpha r^2-\beta r} \quad (11)$$

$$R''(r)=F''(r)e^{-\alpha r^2-\beta r}+F'(r)(-2\alpha r-\beta)e^{-\alpha r^2-\beta r}+\left[(-2\alpha)+(-2\alpha r-\beta)(-2\alpha r-\beta)\right]F(r)e^{-\alpha r^2-\beta r} \quad (12)$$

Substituting Eqs.(10), (11) and (12) into Eq.(6) we have,

$$F''(r)+\left[-4\alpha r-2\beta+\frac{2}{r}\right]F'(r)+\left[\begin{array}{l}(4\alpha^2-J)r^2+(4\alpha\beta-H)r\\+(G-2\beta)\frac{1}{r}-\frac{L(L+1)}{r^2}+(\varepsilon+\beta^2-6\alpha)\end{array}\right]F(r)=0 \quad (13)$$

The function $F(r)$ is considered as a series of the form

$$F(r)=\sum_{n=0}^{\infty}a_n r^{2n+L} \quad (14)$$

Taking the first and second derivatives of Eq.(14) we obtain

$$F'(r)=\sum_{n=0}^{\infty}(2n+L)a_n r^{2n+L-1} \quad (15)$$

$$F''(r)=\sum_{n=0}^{\infty}(2n+L)(2n+L-1)a_n r^{2n+L-2} \quad (16)$$

The substitution of Eqs. (14),(15) and (16) into Eq.(13) gave,

$$\sum_{n=0}^{\infty}(2n+L)(2n+L-1)a_n r^{2n+L-2}+\left[-4\alpha r-2\beta+\frac{2}{r}\right]\sum_{n=0}^{\infty}(2n+L)a_n r^{2n+L-1}$$
$$+\left[(4\alpha^2-J)r^2+(4\alpha\beta-H)r+(G-2\beta)\frac{1}{r}-\frac{L(L+1)}{r^2}+(\varepsilon+\beta^2-6\alpha)\right]\sum_{n=0}^{\infty}a_n r^{2n+L}=0 \quad (17)$$

By collecting powers of $r$ in Eq. (17) we have

$$\sum_{n=0}^{\infty}a_n\left\{\begin{array}{l}[(2n+L)(2n+L-1)+2(2n+L)-L(L+1)]r^{2n+L-2}\\+[-2\beta(2n+L)+(G-2\beta)]r^{2n+L-1}\\+[-4\alpha(2n+L)+\varepsilon+\beta^2-6\alpha]r^{2n+L}\\+[4\alpha\beta-H]r^{2n+L+1}+[4\alpha^2-J]r^{2n+L+2}\end{array}\right\}=0 \quad (18)$$

Equation (18) is linearly independent implying that each of the terms is separately equal to zero, noting that $r$ is a non-zero function; therefore, it is the coefficient of $r$ that is zero. With this in mind, we obtain the relation for each of the terms.

$$(2n+L)(2n+L-1)+2(2n+L)-L(L+1)=0 \quad (19)$$

$$-2\beta(2n+L)+G-2\beta=0 \quad (20)$$

$$-4\alpha(2n+L)+\varepsilon+\beta^2-6\alpha=0 \quad (21)$$

$$4\alpha\beta-H=0 \quad (22)$$

$$4\alpha^2-J=0 \quad (23)$$

From Eq. (20)

$$\beta=\frac{G}{4n+2L+2} \quad (24)$$

From Eq. (23)

$$\alpha = \frac{\sqrt{J}}{2} \quad (25)$$

We proceed to obtaining the energy eigenvalue equation using Eq. (21) and have

$$\varepsilon = 2\alpha(4n + 2L + 3) - \beta^2 \quad (26)$$

Substituting Eqs. (7), (9), (24) and (25) into Eq. (26) and simplifying we obtain

$$E_{nl} = \sqrt{\frac{\hbar^2 \alpha_2}{2\mu}}\left(4n + 2 + \sqrt{(2l+1)^2}\right) - \frac{2\mu\alpha_0^2}{\hbar^2}\left(4n + 1 + \sqrt{(2l+1)^2}\right)^{-2} + \alpha_3 \quad (27)$$

Substituting Eq. (5) into Eq. (27) we obtain the energy eigenvalue equation for quarkonium potential

$$E_{nl} = \sqrt{\frac{\hbar^2}{2\mu}\left(\frac{am_D^3(T)}{6} - \frac{am_D(T)}{3}\right)}\left(4n + 2 + \sqrt{(2l+1)^2}\right)$$
$$- \frac{2\mu}{\hbar^2}\left(a - \frac{2a}{m_D^2(T)} + \frac{2b}{m_D^2(T)}\right)^2\left(4n + 1 + \sqrt{(2l+1)^2}\right)^{-2} + \frac{2b}{m_D(T)} - \frac{2a}{m_D(T)} \quad (28)$$

## RESULTS AND DISCUSSION

The mass spectra of the heavy mesons such as charmonium and bottomonium is calculated using the following relation (Abu-Shady,2016;Inyang et al., 2021)

$$M = 2m + E_{nl}, \quad (29)$$

where $m$ is quarkonium mass, and $E_{nl}$ is energy eigenvalues. By substituting Eq. (28) into Eq. (29) we obtain the mass spectra for quarkonium potential as:

$$M = 2m + \sqrt{\frac{\hbar^2}{2\mu}\left(\frac{am_D^3(T)}{6} - \frac{am_D(T)}{3}\right)}\left(4n + 2 + \sqrt{(2l+1)^2}\right)$$
$$- \frac{2\mu}{\hbar^2}\left(a - \frac{2a}{m_D^2(T)} + \frac{2b}{m_D^2(T)}\right)^2\left(4n + 1 + \sqrt{(2l+1)^2}\right)^{-2} + \frac{2b}{m_D(T)} - \frac{2a}{m_D(T)} \quad (30)$$

We calculate mass spectra of charmonium and bottomonium for quantum states from 1S to 1F using Eq. (30). The free parameters of Eq. (30) were then obtained by solving two algebraic equations.

The experimental data were taken from (Tanabashi et al., 2018). For bottomonium and charmonium systems we adopt the numerical values of these masses as $m_b = 4.823\,GeV$ and $m_c = 1.209\,GeV$ respectively (Barnett et al., 2012). Then, the corresponding reduced mass are $\mu_b = 2.4115\,GeV$ and $\mu_c = 0.6045\,GeV$. The Debye mass $m_D(T)$ is taken as $1.52\,GeV$ by fitted with experimental data. We note that calculation of mass spectra of charmonium and bottomonium are in good agreement with experimental data, as shown in Tables 1 and 2. The values obtained are also in good agreement with work of other researchers like; (Abu-Shady,2016) as shown in Tables 1 and 2 in which the author investigated the N- radial SE analytically. The Cornell potential was extended to finite temperature. In order to test for the accuracy of the predicted results, we used a Chi square function to determine the error between the experimental data and theoretical predicted values. The maximum error in comparison with the experimental data is found to be $0.058\,GeV$.

**Table 1.** Mass spectra of charmonium in (GeV) ($m_c = 1.209$ GeV, $\mu = 0.6045$ GeV, $a = 0.334$ GeV, $b = 0.065$ GeV, $m_D(T) = 1.52$ GeV, $\hbar = 1$)

| State | Present work | NU(Abu-Shady,2016) | AIM(Ciftci & Kisoglu 2018) | Experiment (Tanabashi et al., 2018). |
|---|---|---|---|---|
| 1S | 3.096 | 3.096 | 3.096 | 3.096 |
| 2S | 3.686 | 3.686 | 3.672 | 3.686 |
| 1P | 3.512 | 3.255 | 3.521 | 3.525 |
| 2P | 3.774 | 3.779 | 3.951 | 3.773 |
| 3S | 4.040 | 4.040 | 4.085 | 4.040 |
| 4S | 4.264 | 4.269 | 4.433 | 4.263 |
| 1D | 3.686 | 3.504 | 3.800 | 3.770 |
| 2D | 4.276 | - | - | 4.159 |
| 1F | 3.981 | - | - | - |

**Table 2.** Mass spectra of bottomonium in (GeV) ($m_b$ =4.823 GeV, $\mu$ = 2.4115 GeV, $a$ = 1.152 GeV, $b$ = 0.299 GeV, $m_D(T)$ = 1.52 GeV, $\hbar$ = 1)

| State | Present work | NU(Abu-Shady,2016) | AIM(Ciftci & Kisoglu 2018) | Experiment(Tanabashi *et al.,* 2018). |
|---|---|---|---|---|
| 1S | 9.460 | 9.460 | 9.462 | 9.460 |
| 2S | 10.023 | 10.023 | 10.027 | 10.023 |
| 1P | 9.889 | 9.619 | 9.963 | 9.899 |
| 2P | 10.260 | 10.114 | 10.299 | 10.260 |
| 3S | 10.355 | 10.355 | 10.361 | 10.355 |
| 4S | 10.579 | 10.567 | 10.624 | 10.580 |
| 1D | 10.164 | 9.864 | 10.209 | 10.164 |
| 2D | 10.575 | - | - | - |
| 1F | 10.299 | - | - | - |

## CONCLUSION

In this study, we adopt a quarkonium potential for quark-antiquark interaction. We obtained the approximate solutions of the Schrödinger equation for energy eigenvalues using the series expansion method. The present results were applied to compute heavy-meson masses of charmonium and bottomonium for different quantum states. The result agreed with experimental data and work of other researchers with a maximum error of $0.058\, GeV$.